\documentclass[usenatbib]{mn2e}

\usepackage{pdflscape}
\usepackage{subfigure}
\usepackage{graphicx}
\usepackage{longtable}
\usepackage{float}
\usepackage{multirow}
\usepackage{adjustbox}

\title[Galaxies hosting metal-rich Ly-$\alpha$ absorption systems ]{Magellan LDSS3 emission confirmation of galaxies hosting metal-rich Lyman-$\alpha$ absorption systems}

\author[Lorrie A. Straka]
{Lorrie A. Straka,$^{1}$  Sean Johnson,$^2$ Donald G. York,$^{2,3}$ David V. Bowen,$^4$ Michael Florian,$^2$ \\
\newauthor
Varsha P. Kulkarni,$^5$ Britt Lundgren,$^6$ Celine P\'eroux$^7$\\
$^1$Sterrewacht Leiden, Leiden University, PO Box 9513, NL-2300 RA Leiden, the Netherlands \\
$^2$Department of Astronomy \& Astrophysics, University of Chicago, Chicago, IL 60637, USA\\
$^3$Enrico Fermi Institute, University of Chicago, Chicago, IL 60637, USA\\
$^4$Department of Astrophysical Sciences, Princeton University, Princeton, NJ 08544, USA \\
$^5$Department of Physics \& Astronomy, University of South Carolina, Columbia, SC 29208, USA\\
$^6$Department of Astronomy, University of Wisconsin, Madison, WI 53706, USA\\
$^7$Aix Marseille Universit\'e, CNRS, Laboratoire d'Astrophysique de Marseille, UMR 7326, 13388, Marseille, France\\
}

\date{} 

\pagerange{\pageref{firstpage}--\pageref{lastpage}} \pubyear{}

\def\LaTeX{L\kern-.36em\raise.3ex\hbox{a}\kern-.15em
    T\kern-.1667em\lower.7ex\hbox{E}\kern-.125emX}

\begin{document}

\maketitle

\label{firstpage}

\begin{abstract}

Using the Low Dispersion Survey Spectrograph 3 at the Magellan II Clay Telescope, we target {candidate absorption host galaxies} detected in deep optical imaging {(reaching limiting apparent magnitudes of 23.0-26.5 in $g, r, i,$ and $z$ filters) in the fields of three QSOs, each of which shows the presence of high metallicity, high $N_{\rm HI}$ absorption systems in their spectra (Q0826-2230: $z_{abs}$=0.9110, Q1323-0021: $z_{abs}=0.7160$, Q1436-0051: $z_{abs}=0.7377, 0.9281$). We confirm three host galaxies {at redshifts 0.7387, 0.7401, and 0.9286} for two of the Lyman-$\alpha$ absorption systems (one with two galaxies interacting). For these systems, we are able to determine the star formation rates (SFRs); impact parameters (from previous imaging detections); the velocity shift between the absorption and emission redshifts; and, for one system, also the emission metallicity.} Based on previous photometry, we find these galaxies have L$>$L$^{\ast}$. The [O II] SFRs for these galaxies are in the range $11-25$ M$_{\odot}$ yr$^{-1}$ {(uncorrected for dust)}, while the impact parameters lie in the range $35-54$ kpc.  {Despite the fact that we have confirmed galaxies at 50 kpc from the QSO,  no gradient in metallicity is indicated between the absorption metallicity along the QSO line of sight and the emission line metallicity in the galaxies.} We confirm the anti-correlation between impact parameter and $N_{\rm HI}$ from the literature. We also report the emission redshift of five other galaxies: three at $z_{em}>z_{QSO}$, and two (L$<$L$^{\ast}$) at $z_{em}<z_{QSO}$ not corresponding to any known absorption systems.
\end{abstract}


\section{Introduction}\label{section-introduction}

Quasar absorption line systems have been known and studied for many decades now \citep[e.g.][]{sandage65, york86, noterdaeme09}. These absorption systems are sensitive, redshift independent probes of metals and HI gas down to low densities that are otherwise very difficult to detect. The strongest of these absorption systems are damped Lyman-$\alpha$ systems (DLAs) with neutral hydrogen column densities log $N_{HI}$ $> 20.3$ cm$^{-2}$ and sub-DLAs with $19.0 <$ log $N_{HI}$ $< 20.3$ cm$^{-2}$. These systems dominate the mass density of neutral gas in the universe \citep[e.g.][]{peroux05, prochaska08, zafar13b}, and may be {  galaxy disks or their progenitors}. Tracing the evolution of DLAs and sub-DLAs over cosmic time is therefore critical to understanding galaxy evolution.

With the onset of large spectroscopic surveys such as SDSS \citep{york00}, more than 1,000 DLAs have been detected and compiled in the literature \citep{prochaska05, noterdaeme12b}. While the majority of DLAs are metal-poor, a small fraction of DLAs and a substantial fraction of sub-DLAs appear to be metal-rich, some even super-solar \citep[e.g.][]{kulkarni05, herbertfort06, peroux06a,kulkarni07, meiring09a, som13, som15}. These metal-rich absorbers may form an important link between the largely metal-poor general DLA population and the metal-rich Lyman-break galaxies (LBGs).

In order to understand this relationship, it is vital to study the central 1 kpc of host galaxies and their surrounding IGM to constrain the roles of inflows of pristine gas and outflows of chemically enriched gas in galaxy formation \citep[e.g.][]{dekel09, bouche10, hennawi13}. In addition, it is essential to characterize the stellar content of DLA and sub-DLA host galaxies in order to understand the types of regions in which these metal-rich absorbers arise, such as massive, luminous galaxies or star-burst driven outflows.  However, despite recent progress in detecting their host galaxies \citep[e.g.][]{moller02, peroux11a, peroux12}, fewer than two dozen such systems are confirmed in emission \citep[e.g.][]{rauch08, fynbo10, rauch11, schulze12, krogager12}.

Various techniques have been attempted to obtain this emission confirmation, including Fabry-Perot narrowband imaging \citep{kulkarni06, straka10}. This technique is severely limited by the seeing conditions from the ground due to the small angular separation of the galaxy and quasar. Though many candidate host galaxies have been detected in broad-band imaging, these candidates require followup spectroscopy to confirm whether the emission redshift of the galaxy matches that of the absorber \citep[e.g.][]{chun10, rao11, straka11}. The primary difficulty in detecting emission from DLAs and sub-DLAs is their low {  surface brightness} compared with the background quasars at small angular separations. Therefore, it remains an open question how these absorption features are related to their host galaxies. 

{In recent years,} great progress has been made by observing Mg II, DLA, and sub-DLA systems with integral field spectroscopy. \citet{bouche07} detected 14 out of 21 Mg II absorber host galaxies at $z\sim1$ using SINFONI at VLT. {However, \citet{bouche12} detect only 4 out of 20 Mg II absorber host galaxies at $z\sim2$.} { \citet{peroux11a, peroux12} have confirmed 4 out of 10 Lyman-$\alpha$ absorber host galaxies in H$\alpha$ emission using SINFONI on the VLT at $z\sim1$, but only 1 out of 12 at $z\sim2$. The remaining fields {from this study} have strict limits placed on their star formation rates (SFR) and masses. These HI-selected systems are a subset of the Mg II-selected systems. This 25\% ($z\sim1$: 40\%, $z\sim2$: 8\%) detection rate for Lyman-$\alpha$ host galaxies and 44\% ($z\sim1$: 67\%, $z\sim2$: 20\%) detection rate for Mg II host galaxies in only 2 hours per field is an unprecedented success. However, the number statistics are still low and definitive trends cannot be established based on these few detections. }

We have made imaging detections in previous studies of candidate Lyman-$\alpha$ absorption system host galaxies along the lines of sights to three QSOs using optical and near-infrared broad-band imaging \citep{chun10,straka11,meiring11}.  Here we present spectral data for candidate host galaxies in three of these fields from the LDSS3 spectrograph at the Magellan II Clay Telescope in Chile to contribute to the growing number of successfully studied absorber host galaxies. The paper is arranged as follows. Section~\ref{section-sample} details our sample selection and observations. Section~\ref{section-objects} presents the details of the QSO lines of sight and individual objects in our sample.  Section~\ref{section-results} discusses our results. We present our conclusions and future work in Section~\ref{section-conclusions}. Throughout this paper we have assumed the concordance cosmology: $H_{0}=70$ km s$^{-1}$ Mpc$^{-1}$, $\Omega_{M}=0.30$, and $\Omega_{\Lambda}=0.70$. 

\begin{table*}
\centering
\resizebox{1.0\textwidth}{!}{
\begin{minipage}{175mm}
\caption{Target list\label{tbl-targets}}
\begin{tabular*}{\columnwidth}{@{\extracolsep{\stretch{1}}}*{11}{lccccccccccc}@{}}

\hline\hline

Object & RA & Dec & z$_{QSO}$ & z$_{abs}$ & log N$_{HI}$ & [Zn/H]$^1$  & W$^{rest}_{\lambda2796}$ & Ref\\
 & & & & & (cm$^{-2}$) & & (\AA) & (N(HI), [Zn/H], W)  \\

\hline

Q0826-2230 & 08:26:01.5 & -22:30:26.2 &  $>0.911$ & 0.9110 & 19.04$\pm0.04$ & $+$0.68  & 1.15$\pm0.2$ & (1;2;4)  \\
Q1323-0021 & 13:23:24.3 & -00:22:30.9 & 1.388 & 0.7160 & 20.54$\pm0.15$ & $+$0.61  & 2.41$\pm0.07$ & (1;2;3) \\
Q1436-0051 & 14:36:45.05 & -00:51:50.61 & 1.28 & 0.7377 & 20.08$\pm0.10$ & -0.05  & 1.14$\pm0.08$ & (1;2;3) \\
	& 14:36:45.05 & -00:51:50.61 & 1.28 & 0.9281 & $18.4\pm0.98$ & $-0.05\pm0.55$ & 1.20$\pm0.07$ & (5;2,5;1)\\

\hline

\end{tabular*}

$^1${\cite{rao06}}
$^2${\cite{meiring09b}}
$^3${\cite{nestor05}}
$^4${\cite{falomo90}}
$^5${This work.}

\end{minipage}
}
\end{table*}

\section{Observations}\label{section-sample}

\subsection{Sample Selection}

We have chosen three QSO lines of sight with metal-rich, high N$_{\rm HI}$ absorption systems. Q0826-2230, Q1323-0021, and Q1436-0051 all show super-solar absorption systems ([Zn/H]\footnote{The metallicity is defined by  [X/H] = log(N$_X$/N$_H$)$_{obs}$ - log(N$_X$/N$_H$)$_{\odot}$.}$>-0.05$) with strong equivalent width neutral hydrogen absorption systems (Q0826-2230: log $N_{\rm HI}=19.04$, Q1323-0021: log $N_{\rm HI}=$20.54, Q1436-0051: log $N_{\rm HI}=$20.08 and $17.45<$log N$_{\rm HI}<18.8$).  Q1436-0051 has two absorption systems. We target this field primarily for the sub-DLA with log $N_{\rm HI}=20.08$ with the added benefit of searching for the host galaxy of this additional Lyman Limit System. We use zinc as a tracer of metallicity due to its undepleted nature in the interstellar medium \citep[e.g.][]{york82b, meyer89, pettini90}.  { Table~\ref{tbl-targets} lists our QSO targets and their previously observed absorption properties, which were taken from the literature (RA, Dec, z$_{QSO}$, z$_{abs}$, log N$_{HI}$, [Zn/H], and W$^{rest}_{\lambda2796}$), along with their literature references. }

\subsection{LDSS3 Data}

We have obtained data using the Low Dispersion Survey Spectrograph 3 (LDSS3) located on the Magellan II Clay Telescope in Chile. LDSS3 is a multi-object spectrograph which {  can be used to obtain large numbers of faint objects simultaneously over a field of view of $8.3\arcmin$ in diameter}. Each of our three fields was observed between March and April 2011 with the VPH-Red grism. Exposure times were 1600-1800 seconds, with 2 exposures for two fields and 4 exposures for one field. Seeing ranged from 0.6\arcsec to 0.8\arcsec. Table~\ref{tbl-observations} lists our journal of observations.

All targets have been observed using the VPH-Red grism, which covers a spectral range of 6000 \AA ~to 10000 \AA ~with a resolution of  R$\sim$1810. This spectral range allows us to obtain the strongest lines available in the optical spectrograph range at the redshift of the observed neutral hydrogen absorption systems. These lines include primarily [O II] $\lambda3727$, H$\beta$, and [O III] $\lambda\lambda4960, 5008$.  H$\alpha$ and [N~II] are redshifted into the NIR at the redshift of these absorption systems. The LDSS3 data were reduced using the COSMOS (Carnegie Observatories System for Multi-Object Spectroscopy) reduction software \footnote{http://code.obs.carnegiescience.edu/cosmos}. We have flux calibrated the spectra against the QSO in the field (and interpolating over the positions of the QSO emission lines). 

Detections of [O II] allow us to estimate the SFR following the prescription of \citet{kennicutt98}:

\begin{equation}
SFR_{[O II]}~(M_{\sun}~yr^{-1}) = 1.4\times10^{-41}~L([O II]) (ergs ~s^{-1})
\end{equation}
where L([O II]) is the luminosity in [O II] emission. {Without NIR spectral observations of these targets to measure H$\alpha$, we  cannot estimate the Balmer decrement reddening values. Such observations would allow us to determine the role of dust in the environment. Therefore, the SFRs are not corrected for dust. We note that without this correction, the [O II] SFR shows significant scatter (up to 32\% even for corrected values) and should be taken as a cautious lower limit \citep{moustakas06}. Therefore the uncertainties on these SFR values are $\sim 0.39$ dex.} 


\begin{table}
\centering
\resizebox{1.0\linewidth}{!}{
\begin{minipage}{1.0\linewidth}
\caption{Journal of observations\label{tbl-observations}}
\begin{tabular*}{\columnwidth}{@{\extracolsep{\stretch{1}}}*{5}{lcccc}@{}}

\hline\hline

Object & Date & Exp. (s) & No. & Seeing (\arcsec)  \\

\hline

Q0826-2230 & Apr 2011 & 1800 & 4   & 0.6-0.8 \\
Q1323-0021 & Apr 2011 & 1800 & 2  & 0.7-0.8 \\
Q1436-0051 & Mar 2011 & 1600 & 2   & 0.6-0.7 \\

\hline

\end{tabular*}

\end{minipage}
}
\end{table}

\subsection{Photometry}

For each of our targets with confirmed redshifts, we calculate the absolute i-band magnitude and L$_{i}$/L$_{i}^{\ast}$, based on the apparent i-band magnitudes in \citet{meiring11}. For targets without confirmed redshifts, we calculate the limiting absolute i-band magnitude and L$_{i}$/L$_{i}^{\ast}$ for the field at the redshift of the absorption system. We implement k-corrections for three spectral types: E, Sa, and burst, using the external IRAF package COSMOPACK \citep{balcells03}. Table~\ref{tbl-photometry} details these measurements, including the QSO; object ID; $z_{em}$; impact parameter (b); r-band apparent magnitude; i-band absolute magnitude for E, Sa, and burst; and L$_{i}$/L$_{i}^{\ast}$ for E, Sa, and burst. Details of these limits are discussed in the sections on the individual fields. We refer the reader to \citet{meiring11} for an in depth analysis of the imaging data and photometry.

\section{Individual Objects}\label{section-objects}

\begin{figure*}

\subfigure[Q0826-2230]{\label{fig-q0826_soar}
\includegraphics[trim = 0mm 0mm 0mm 0mm, clip, width=0.3\textwidth]{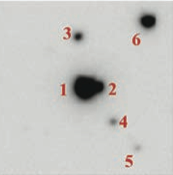}
 }
 \subfigure[Q1323-0021]{\label{fig-q1323_soar}
 \includegraphics[trim = 0mm 0mm 0mm 0mm, clip, width=0.31\textwidth]{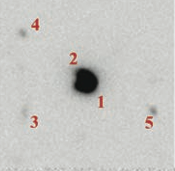}
 }
  \subfigure[Q1436-0051]{\label{fig-q1436_soar}
 \includegraphics[trim = 0mm 0mm 0mm 0mm, clip, width=0.3\textwidth]{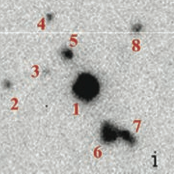}
 }
 
\caption{Deep i-band images from the SOAR optical imager in \citet{meiring11}. Each thumbnail is 20 arcsec $\times$ 20 arcsec. North is up and east is left. \label{fig-soar_images} }
 
\end{figure*}

Here we present the results of our spectroscopic study. We confirm the host galaxies for two absorption systems out of four in three fields. All objects are numbered according to the scheme of \citet{meiring11}. {  Figure~\ref{fig-soar_images} shows the SOAR i-band images for each of the fields in our sample from \citet{meiring11}.} We detail our detections below. For brevity, we do not discuss the objects from Meiring et al. that were not targeted here in our spectroscopic study. They are still labeled in Figure~\ref{fig-soar_images} for reference. Table~\ref{tbl-characteristics} lists the emission line measurements from our data: QSO; object ID; impact parameter; $r$-band apparent magnitude; relative velocity shift between the absorption systems and detected host galaxies; emission redshift; [O II], H$\beta$, [O III], and H$\alpha$ fluxes; [O II] or H$\alpha$ SFR, and emission metallicity (12+log(O/H)).  

\subsection{Q0826-2230}

Q0826-2230 has a sub-DLA with log N$_{\rm HI}$ $=19.04\pm0.04$ \citep{rao06} at $z_{abs}=0.9110$ and metallicity [Zn/H]$=+0.68\pm0.08$ \citep{meiring07}.  \cite{meiring11}  report optical imaging detections for 6 objects in this field.   Figure~\ref{fig-q0826_soar} shows the SOAR image for this field from \citet{meiring11} and labels each of the galaxies targeted in their study. { The limiting absolute magnitude and corresponding L$_{i}$ value for this field at the redshift of the absorber are M$_{i}>-20.36$ and L$_{i}<0.36$L$^{\ast}_{i}$ (for Sa type galaxies). 

We are limited in the number of galaxies we can target by the slit placement in the LDSS3 slit masks. For the field of Q0826-2230, we obtain spectra for objects 3, 4, and 5, plus one additional serendipitous object not in the Meiring et al. field of view which we label object 7.  Not targeted here are objects 2 and 6. However, it is suggested that this QSO is gravitationally lensed, and object 2 is an image of the QSO \citep[e.g.][]{falomo90, meiring11}. Object 6 is a star based on imaging analysis.  We spectroscopically confirm object 3 as a star.}

{Despite the fact that the image of Meiring et al. covers impact parameters far exceeding 100 kpc for objects at $z=0.9110$, no emission line galaxies at this redshift were discovered. This is consistent with the findings of \citet{bouche12} and \citet{peroux12} (see above) that Ly$\alpha$ and Mg II absorbers are not always accompanied by detectable emission galaxies.} This indicates that the host galaxy is either very faint (fainter than the limits reached by \cite{meiring11} at this redshift: M$_{i, E}>-20.63$, M$_{i, Sa}>-20.36$) or falls within the PSF of the QSO and cannot be resolved (likely both). {  

Three objects did show up at different redshifts in this field ($z\sim 0.1070, 0.9512, 1.1457$), determined by a single emission line in each case. Since absorbers at the possible redshifts of these emission lines were not found in the spectrum of Q0826-2230, confirmation of the emission line redshifts (see below) would confirm that these are examples of emission line galaxies without detectable Mg II absorption at reasonable impact parameters.} 

Objects 4 and 5 both appear to have emission at $\sim$8000 \AA~ in the 2D spectrum. However, the line is not significant in the 1D spectra extraction. This single line detection can be either [O II] or H$\alpha$. If [O II], it corresponds to a redshift of $z_{em}=1.1457$, but if H$\alpha$ it corresponds to $z_{em}=0.2186$. Spectral coverage below 5800 \AA~  or above 9500 \AA ~is necessary to confirm this line. At these redshifts, these galaxies would have L$_{i}$ values of 36.6L$^{\ast}_{i}$ and 3.6L$^{\ast}_{i}$, respectively. The high redshift of these systems likely results in an inaccurate k-correction. 

Object 7 in Table~\ref{tbl-targets} is not in the field of view of consideration for the study from \cite{meiring11}, but has strong emission which could be either H$\alpha$ at $z_{em}=0.1070$ or [O II] at $z_{em}=0.9512$. If it is the latter, it is at a significant offset from the absorption redshift (6200 km s$^{-1}$), and so we still do not consider it a candidate for the host galaxy.  This emission line is strong and broad (441 km s$^{-1}$ at FWHM), which is below the typical line width definition of a Seyfert galaxy ($>500$ km s$^{-1}$), and so this line is likely the [O II] doublet unresolved. 

\begin{figure}

\includegraphics[trim = 15mm 10mm 10mm 90mm, clip, width=0.5\textwidth]{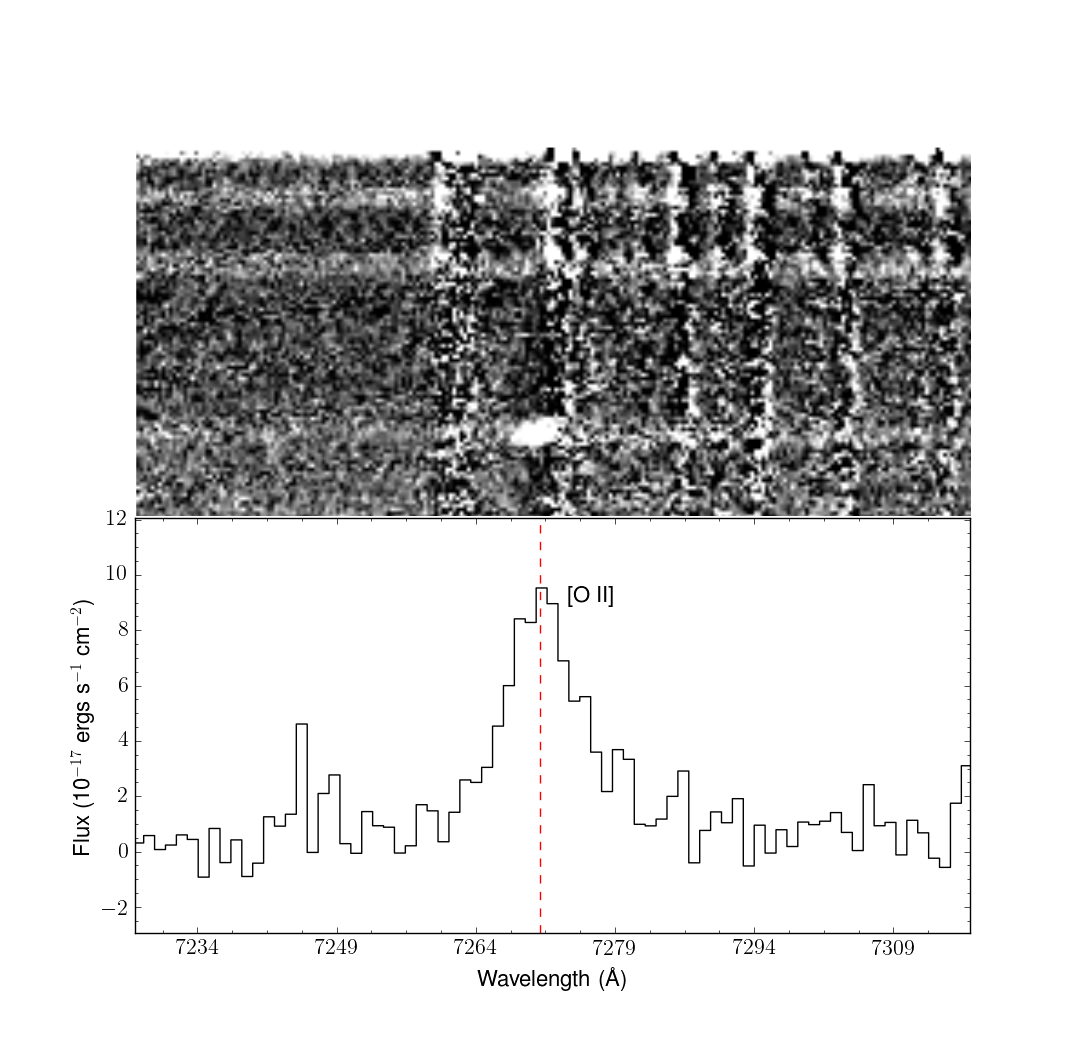}
 
\caption{Q0826-2230 Object 7. This emission line corresponds to a redshift of $z=0.9512$ for [O II], but is not in the field of view of \citet{meiring11} and so does not have a corresponding object number or imaging detection. The [O II] emission corresponds to a SFR of 57.3 M$_{\odot}$ yr$^{-1}$.  { Even considering doublet separation, this line is very broad (441 km s$^{-1}$ at FWHM) and is possibly a Seyfert galaxy.} \label{fig-q0826_45_3}}
 
\end{figure}

{Based on SOAR archival data, we are able to determine the impact parameter of this object with the QSO line of sight. At $z=0.1070$, object 7 is at an impact parameter of b$=35.5$ kpc. At $z=0.9512$, it is at b$=143.2$ kpc. Q0826-2230 is a BL Lac object, and its redshift remains unknown, except that it must be above the absorption redshift of the sub-DLA. Therefore, the QSO could be beyond $z=0.9512$, in which case the absence of absorption at this redshift \citep{meiring07} is interesting. The limit on Ca II H and K absorption at $z=0.9512$ is EW$<0.01$ \AA. There is also no known absorption system at $z=0.1070$, with a Mg II EW limit of $0.02$ \AA. } Figure~\ref{fig-q0826_45_3} shows the 2D and 1D emission for this object.


\subsection{Q1323-0021}

Q1323-0021 has a sub-DLA with log N$_{\rm HI}$ $=20.21\pm0.20$ at $z_{abs}=0.7160$ \citep{peroux06a} and metallicity [Zn/H]$=+0.61\pm0.20$ \citep{khare04, peroux06a}. \citet{chun10} use the Keck Laser Guide Star Adaptive Optics (LGSAO) to obtain deep K'-band imaging of this field and succeeded in detecting a galaxy at an impact parameter of 9 kpc \citep[object 2]{meiring11}. Though it is possible to see the very edge of this galaxy along the edge of the QSO without AO, and in fact this galaxy is detected in non-AO imaging of \citet{hewett07} and \citet{meiring11} after PSF subtraction of the QSO, its close proximity to the QSO makes it difficult to obtain an emission spectrum. We have placed a slit mask on this galaxy, but the resulting spectrum is dominated by QSO continuum. Further spectroscopic followup with AO is necessary.  However, we note that \citet{peroux11b} target this object with the SINFONI integral field unit (which is not sensitive to continuum) at the Very Large Telescope (VLT) for observations of H$\alpha$ and find no emission down to F$_{H\alpha}<1\times10^{-17}$ ergs cm$^{-2}$ s$^{-1}$, corresponding to a SFR$_{H\alpha}<0.1$ M$_{\odot}$ yr$^{-1}$. These results support the hypothesis that this is an early type galaxy. 

Again, we are limited by the slit placement for LDSS3. For the field of Q1323-0021, we target objects 2, 3, 4, and 5, which are detailed below. Figure~\ref{fig-q1323_soar} shows the SOAR optical i-band image for this field.  { The limiting absolute magnitude and corresponding L$_{i}$ value for this field at the redshift of the absorber are M$_{i}>-18.78$ and L$_{i}<0.09$L$^{\ast}_{i}$ (for Sa type galaxies). }

Object 3 has a very faint continuum trace with no sign of emission. Our data may be too shallow to detect emission from this object, or it may be an early type galaxy with no detectable star formation. { Assuming object 3 is at the redshift of the absorber, the SFR limit from our LDSS3 data is $<0.55$ M$_{\odot}$ yr$^{-1}$.  As with object 2, this is consistent with the much stricter limit SFR $< 0.1$ M$_{\odot}$ yr$^{-1}$ from \citet{peroux11a}. These findings are also consistent with the suggestion that the absorber host galaxy is early-type \citep{chun10,peroux11a}, and that metal-rich sub-DLAs may arise in early-type galaxies \citep[e.g.][]{khare07}. } 

\begin{figure}
 
\includegraphics[trim = 15mm 5mm 10mm 70mm, clip, width=0.5\textwidth]{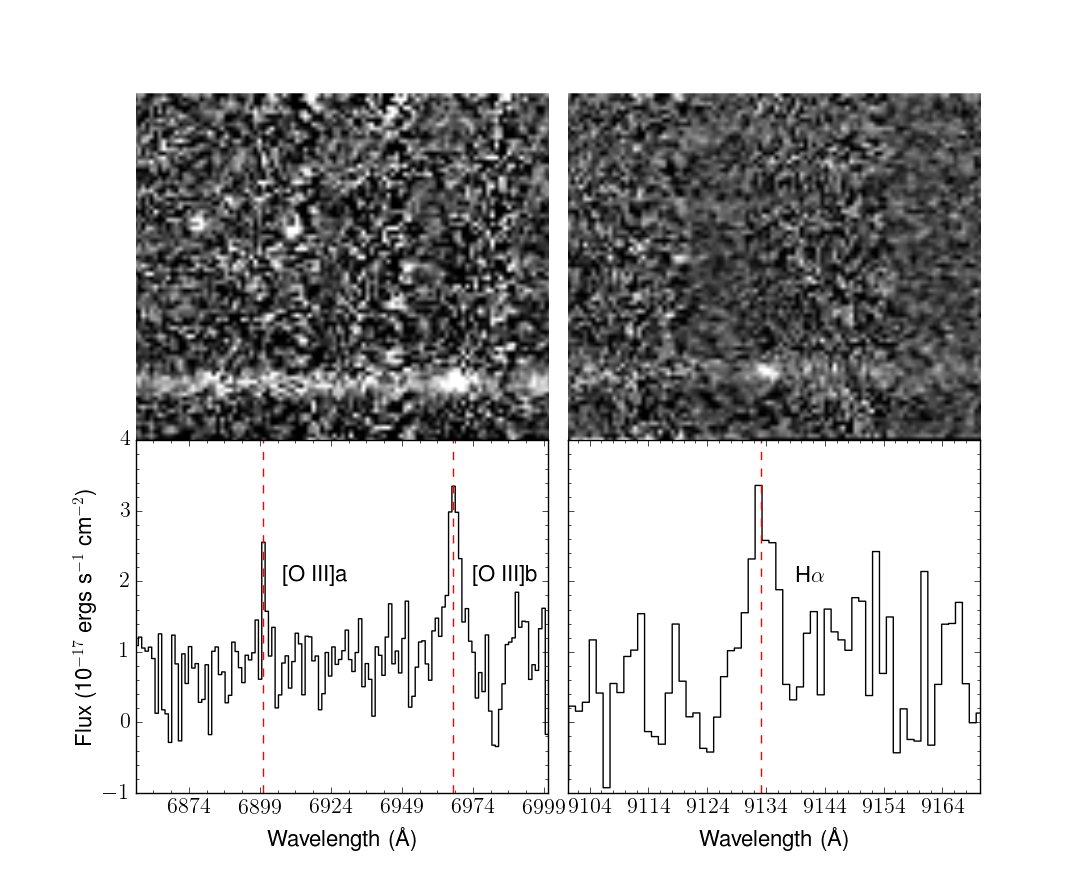}
 
\caption{Q1323-0026 Object 4. Detections of [O III] and H$\alpha$ indicate this galaxy is at a redshift of $z=0.3915$. This corresponds to an impact parameter with the QSO line of sight of 49 kpc.  The H$\alpha$ SFR is 0.69 M$_{\odot}$ yr$^{-1}$.  This galaxy does not correspond to any known absorption system.  \label{fig-q1323_obj4}}
 
\end{figure}

We detect emission for object 4 at $z_{em}=0.3910$ in [O~III] and H$\alpha$ at an impact parameter of b$=49$ kpc. This does not correspond to any known absorption system along the line of sight, as Mg II would be visible in the SDSS spectrum at this redshift and is not detected. We place a $3\sigma$ limit of 0.67 \AA ~on the Mg II observed equivalent width for this system. We determine a SFR of 0.69 M$_{\odot}$ yr$^{-1}$ from the H$\alpha$ emission, based on the adopted \citet{kennicutt98} relation. This is uncorrected for extinction using the Balmer decrement because deeper data are necessary to detect H$\beta$. We have placed an upper limit on the H$\beta$  flux, based on the noise. [O II] at this redshift is at $5184$ \AA, which is below the bluest wavelength of this study.  The corresponding L$_{i}$ is 0.08$\pm0.05$L$^{\ast}_{i}$, based on the i-band apparent magnitude determined by \citet{meiring11}. Figure~\ref{fig-q1323_obj4} shows the 2D spectra and the extracted emission lines for this object. 

\begin{figure}
 
\includegraphics[trim = 10mm 10mm 10mm 120mm, clip, width=0.4\textwidth]{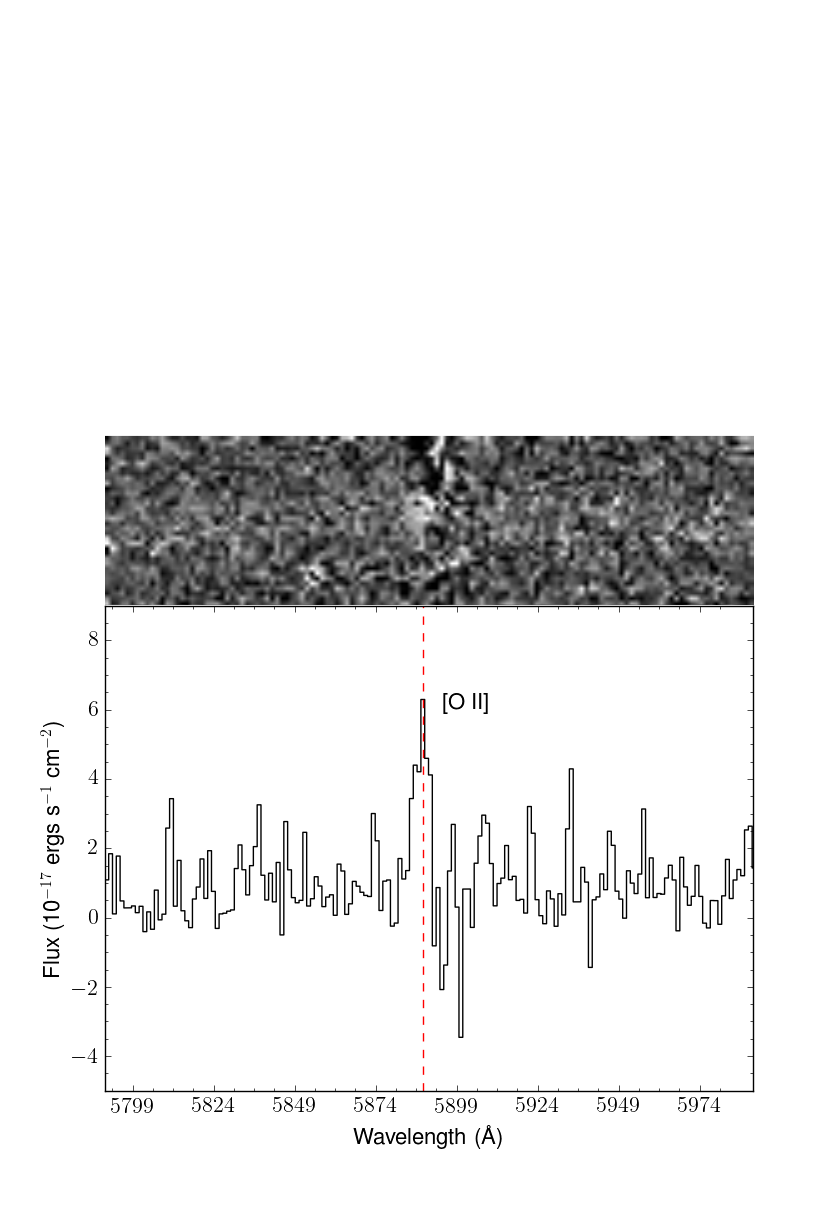}
 
\caption{Q1323-0026 Object 5 at $z=0.5797$. This corresponds to an impact parameter of 55 kpc. The [O II] SFR is 15.1 M$_{\odot}$ yr$^{-1}$.  This galaxy does not corresponds to any known absorption system.  \label{fig-q1323_obj5}}
 
\end{figure}

Object 5 has faint emission at $z_{em}=0.5797$ in [O II]. This corresponds to a SFR of 15.1 M$_{\odot}$ yr$^{-1}$ and impact parameter of b$=55$ kpc. H$\beta$ and the two [O III] lines are lost in sky residuals. NIR data are necessary to determine the flux in H$\alpha$. {As with object 4, there is no detected absorption in the SDSS spectrum at this redshift. We place a $3\sigma$ limit of 0.37 \AA ~on the Mg II observed equivalent width for this system.} The  L$_{i}$ value for this galaxy is 0.17$\pm 0.05$L$^{\ast}_{i}$. Figure~\ref{fig-q1323_obj5} shows the 2D spectra and the extracted emission lines for this object. 

\subsection{Q1436-0051}

Q1436-0051 has two absorption systems along the line of sight: we target this field primarily for a sub-DLA with log N$_{\rm HI}$ $=20.08\pm0.11$ at $z_{abs}=0.7377$ and metallicity [Zn/H]$=-0.05\pm0.12$, but there is also a strong neutral hydrogen absorption system at $z_{abs}=0.9281$ with large errors on the column density.  For this absorber, \citet{rao06} report an upper limit of log N$_{\rm HI}<18.8$ derived from HST STIS data (ID 9382, PI: Rao). \citet{ribaudo11} analyze the STIS archival data via the optical depth of the absorption features,  placing a lower limit of log N$_{\rm HI}> 17.45$. The super-solar metallicity of [Zn/H]$>+0.86$ reported in \citet{meiring08} is derived based on the upper limit reported in \citet{rao06}.  

We analyze the STIS archival data for the absorber at $z=0.9281$ in order to better understand the HI column density and Zn metallicity for our study. Using the Lyman-limit to determine N HI, we determine log N$_{\rm HI}=17.6\pm1.1$. However, the Lyman limit is not robustly detected for this system ($\tau=2.5\pm1.7$) due to the low SNR at the Lyman limit in the STIS QSO spectrum, which drives the large uncertainty. By measuring the equivalent width of the Lyman-$\alpha$ absorption, we determine a column density of 19.4. However, this is based on a single component fit, which is inaccurate. The high resolution spectra for this system shows multiple Mg II components which may contribute significantly to the Lyman-$\alpha$ equivalent width. We do not have enough constraints to accurately decompose the HI absorption into its components. Therefore, we adopt $17.45<$log$N_{\rm HI}<19.4$ as the range of column densities for this system. For purposes of plotting, we adopt log $N_{\rm HI}=18.4\pm0.98$, corresponding to the midpoint of the range with the endpoints representing the error bars.

Based on this new column density range, we have re-calculated the Zn metallicity, the error of which reflects the large range in the column density. The corresponding metallicity range is $-0.60<$[Zn/H]$<+0.50$, after applying appropriate ionization corrections using CLOUDY \citep{ferland13}. A [Zn/H] of -0.60 corresponds to the highest possible HI column density, while $+0.50$ corresponds to the lowest possible HI column density. For the purposes of plotting, we adopt [Zn/H]$=-0.05\pm0.55$, which correspond to the midpoint in the range with the errors corresponding to the endpoints (as with the column density). 

The field of Q1436-0051 has previous detections in broad-band imaging in the NIR \citep{straka11} and  optical \citep{meiring11}. We have focused on obtaining spectra for objects 2, 5, 6, and 7 of \citet{meiring11}. {Figure~\ref{fig-q1436_soar} shows the SOAR i-band image for this field}. It was not possible to obtain simultaneous spectra for all objects in the field due to the close proximity of the slit masks in the field. We chose to obtain spectra for objects 2, 5, 6, and 7. \citet{meiring11} judge the objects we have targeted to be the best candidates for hosting the absorption systems based on $u-r$ color and photometric redshift estimates. 

Object 2 had a very faint trace with no discernible emission, indicating it may be an early type galaxy with little current star formation. { However, the detections of multiple galaxies with perturbations at the same redshift in this field (objects 6 and 7 at $z\sim0.7377$, and possibly object 5 at $z\sim0.9286$) indicate we may be dealing with two different galaxy group environments and therefore we cannot rule out the possibility that object 2 is a member of one of these groups. Table~\ref{tbl-photometry} indicates the possible photometric properties of this galaxy if it were at one of these two redshifts.}

\begin{figure}
 
\includegraphics[trim = 25mm 10mm 0mm 120mm, clip, width=0.40\textwidth]{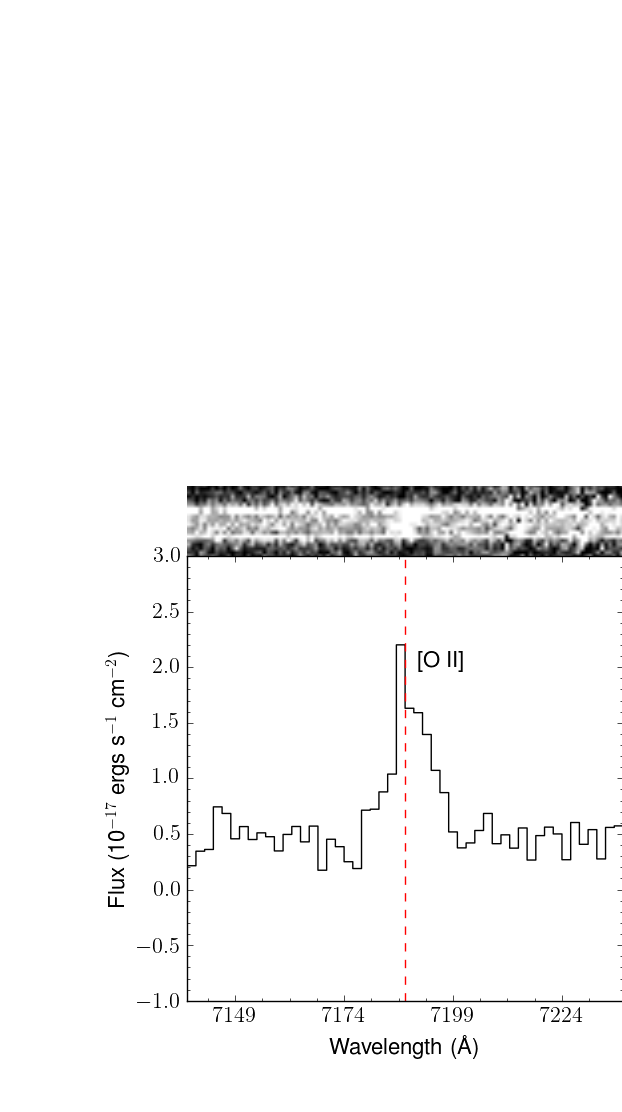}
 
\caption{Q1436-0051 Object 5. The top panel shows the 2D spectrum zoomed in on the emission line. The bottom panel shows the extracted flux calibrated emission line. No other nebular emission lines are present in this spectra, and in conjunction with the asymmetric profile of this line indicates this is indeed [O II] at $z_{em}=0.9286$.  This corresponds to an impact parameter of 35 kpc. The [O II] SFR is 11.6 M$_{\odot}$ yr$^{-1}$.  \label{fig-q1436_5}}
 
\end{figure}

We confirm that object 5 is at $z_{em}=0.9286$, corresponding closely to the redshift of the strong absorber at $z_{abs}=0.9281$ at an impact parameter of b$=35$ kpc. We measure this redshift based on unresolved [O II]$\lambda 3727$. The corresponding relative velocity shift between the centroids of the absorption and emission lines is $-78$ km s$^{-1}$ (given by $\Delta v/c=(z_{abs}-z_{gal})/(1+z_{gal})$). Figure~\ref{fig-q1436_5} shows the 2D spectra and the extracted [O II] emission line. The flux in this line indicates a SFR of 11 M$_{\odot}$ yr$^{-1}$. The corresponding L$_{i}$ is 3.4L$^{\ast}_{i}$. { The image from \citet{meiring11} in Figure~\ref{fig-q1436_soar} shows that this galaxy may have a disturbed morphology due to interaction with a second nearby galaxy. }

\begin{figure*}
 
\includegraphics[trim = 0mm 0mm 0mm 50mm, clip,width=1.0\textwidth]{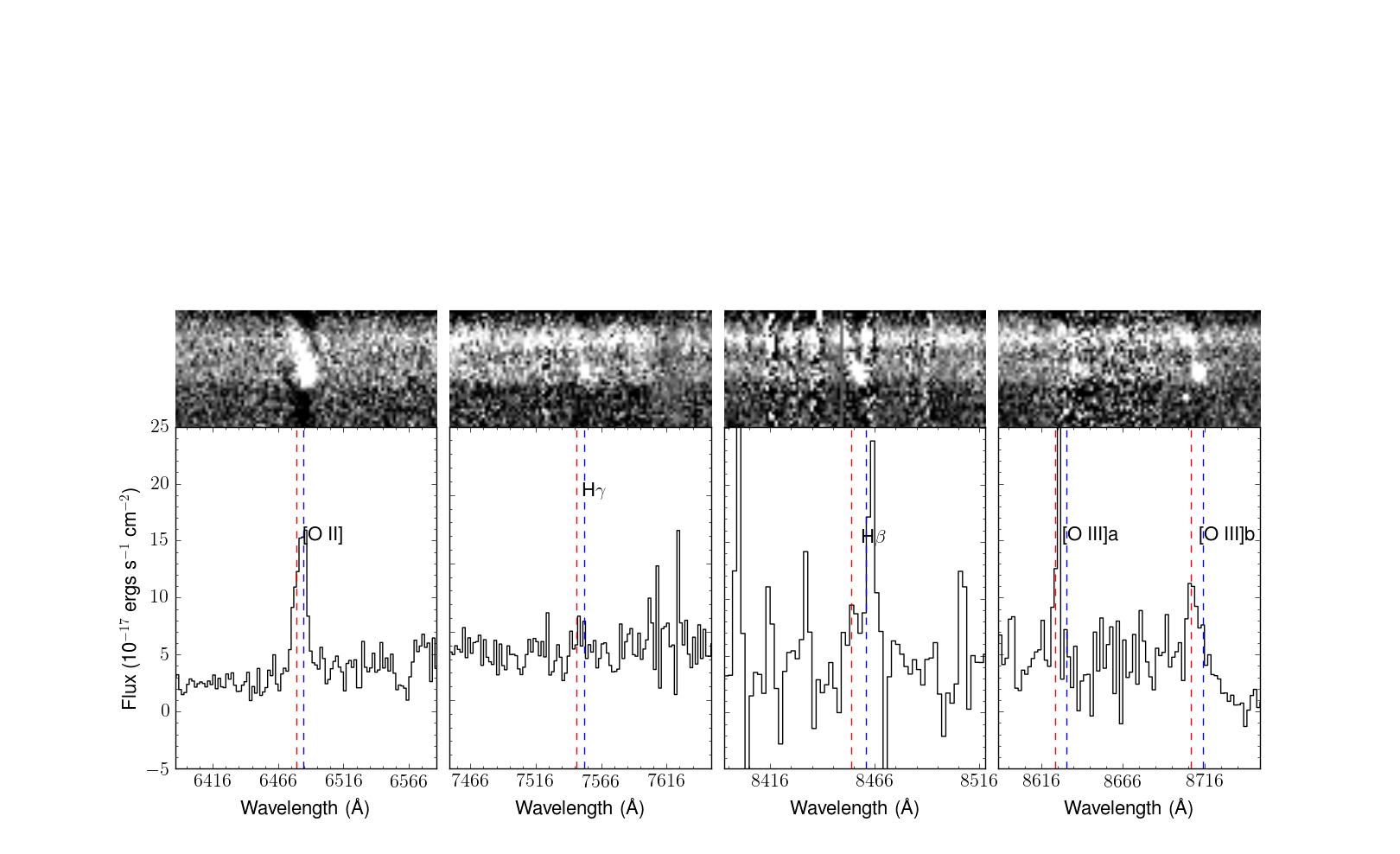}

\caption{Q1436-0051 Objects 6 and 7. The bottom panels show the extracted emission lines for the combined objects (unable to be resolved), while the top panels show the 2D spectra. We detect strong [O II], H$\beta$, and [O III] in both galaxies at redshifts of $z=0.7401$ and 0.7387 respectively. There is a trace of H$\gamma$, but it is only present at less than 1$\sigma$. There is  clear offset between the two galaxies seen in the 2D spectra from which we estimate the redshift difference. These galaxies are located at impact parameters of b$=45$ kpc and 54 kpc respectively. The corresponding combined [O II] SFR is 47.45 M$_{\odot}$ yr$^{-1}$, with estimated individual SFR corresponding to $\sim26$ and $\sim22$ M$_{\odot}$ yr$^{-1}$ respectively.  \label{fig-q1436_67}}

\end{figure*}

We can also confirm that objects 6 and 7 are at $z_{em}=0.7401$ and $z_{em}=0.7387$ (at impact parameters of b=$45$ and 54 kpc) respectively, corresponding to the redshift of the sub-DLA, based on [O II], H$\beta$, and [O III] emission. We calculate L$_{i}$  for these two galaxies based on the i-band apparent magnitudes in \citet{meiring11} and find values of 4.0L$^{\ast}_{i}$ and 1.6L$^{\ast}_{i}$, respectively. The relative velocity shifts between the absorption system ($z_{abs}=0.7377$) and these two galaxies are $-414$ and $-173$ km s$^{-1}$, respectively. These values are consistent with the tilt of the emission lines (evident in Figure~\ref{fig-q1436_67}). We do not resolve the two traces completely, and so we measure the emission of the two galaxies together and then estimate the flux contribution from each from the 2D spectra and broad-band images. The 2D spectra show a clear offset in redshift between the two galaxies, so we use this to determine a more accurate redshift for the individual galaxies. There is a relative offset of $\Delta v= 259$ km s$^{-1}$ between the two. 

These galaxies are actively star forming as posited by \citet{meiring11} due to their relatively flat spectral energy distributions. Together, these two absorber galaxies have a SFR of 48 M$_{\odot}$ yr$^{-1}$ based on [O II] emission. Using the $r-$ and $i-$band magnitudes determined in \citet{meiring11}, we are able to estimate the ratio of total flux belonging to each galaxy. This indicates that in the $r-$band the flux from object 7 is 0.47 that of object 6, and in the $i-$band it is 0.40 that of object 6. Therefore, given the SFR-stellar mass relation \citep[e.g.][]{daddi07}, we conservatively estimate the SFR from objects 6 and 7 to be $\sim26$ M$_{\odot}$ yr$^{-1}$ and  $\sim22$ M$_{\odot}$ yr$^{-1}$, respectively.  Figure~\ref{fig-q1436_67} shows the 2D spectra and the emission lines detected for these objects (extracted together). 

For these objects we can estimate the R$_{23}$ emission metallicity following the prescription of \citet{pagel79}: 

\begin{equation}
R23 = \frac{[OII]\lambda3727 + [OIII]\lambda\lambda 4959,5007}{H\beta}
\end{equation}
and
\begin{equation}
O32=\frac{[OIII]\lambda\lambda4959,5007}{[OII]\lambda3727}
\end{equation}
R$_{23}$ values are then converted to $12+log(O/H)$ lower and upper values according to \citet{kobulnicky99} and \citet{mcgaugh91}, using the O32 indicator.

We estimate that objects 6 and 7 together have an R23 = 0.17$\pm0.10$. Using the flux ratios above, we estimate individual R23 values of R23$_{6}=0.14\pm 0.13$ and R23$_{7}=0.20\pm 0.16$. The solar value is adopted to be $12+log(O/H)_{\odot}=8.69$ \citep{asplund09}. Which branch of the R$_{23}$ diagram these objects fall into is determined by the [O III]$\lambda5007$/[N II]$\lambda6584$ ratio \citep{kobulnicky99}. This combined R23 corresponds to $12+log(O/H)_{l}=7.19\pm0.30$ and $12+log(O/H)_{u}=9.02\pm0.30$, with individual values within 0.05 of the combined lower values and within 0.01 of the combined upper values. NIR spectra are required to determine the [NII] flux in these objects in order to break the degeneracy in R$_{23}$. 

Assuming these objects follow the upper branch of R$_{23}$, the above 12+log(O/H) upper value corresponds to $+0.33$ dex. We estimate the metallicity gradient to be $-0.01\pm0.006$ dex kpc$^{-1}$ between the absorption along the line of sight to the QSO and the galaxy emission. Our finding is consistent with that of \citet{peroux12}, who reported a mixture of mild positive as well as negative metallicity gradients for a sample of three sub-DLAs and six DLAs. { Note that the emission metallicity and the absorption metallicity are not measured in the same phase of the gas (HII and HI, respectively). These gradients illustrate the metallicity with respect to solar as a function of impact parameter, which can be used as an indicator of inflows and outflows of gas. Inverted gradients are reported by \citet{cresci10} as a signature of primordial gas infall. \citet{queyrel12} suggest that this gas infall may be due to interaction or cold gas accretion; in this case, objects 6 and 7 are interacting. However, the gradient measured here is virtually nonexistent, so we rule out infalling gas. The high metallicity absorbing gas may be the result of stripped material due to the interaction between objects 6 and 7, starburst driven outflows, or a combination of stripped and outflowing material. We discuss these scenarios below.}

\begin{table*}
\centering
\resizebox{1.0\textwidth}{!}
{
\begin{minipage}{1.0\textwidth}
\caption{Object photometry \label{tbl-photometry}}
\begin{tabular*}{\columnwidth}{@{\extracolsep{\stretch{1}}}*{11}{llccccccccc}@{}}
\hline\hline
QSO & ID$^1$ & $z_{em}$ & b$^2$ & m$_{gal}^{r3}$  & M$_{i}$ & M$_{i}$ & M$_{i}$ & L$_{i}$/L$_{i}^{\ast}$ & L$_{i}$/L$_{i}^{\ast}$ & L$_{i}$/L$_{i}^{\ast}$   \\ 
& & & (kpc) & & E & Sa & Burst & E & Sa & Burst \\
\hline
Q0826-2230  & limit$^7$ & 0.9110 & -- & -- & $>-20.63$ & $>-20.36$ & $>-19.14$ & $<0.46$ & $<0.36$ & $<0.12$ \\
	& 4 & 1.1457 & 40.9 & 21.44 & -25.75 & -25.37 & -23.68 & 52.2 & 36.6 & 7.7 \\
 	 & 5 & 1.1457 & 77.1 & 23.46 & -24.68 & -22.86 & -23.07 & 19.4 & 3.6 & 4.4 \\
	 & 7 & 0.1070 & 35.5 & -- & -- & -- & -- & -- & -- & --  \\
	 & 7 & 0.9512 & 143.2 & -- &-- & -- & -- & -- & -- & -- \\ \\
Q1323-0021  & limit$^7$ & 0.7160 & -- & -- & $>-18.97$ & $>-18.78$ & $>-17.95$ & $<0.10$ & $<0.09$ & $<0.04$ \\
	 & 2 & --  & 9.0$^4$ & -- &  -- & -- & -- & -- & -- & -- \\
	 & 3 & -- & 58.4 & 23.78 & -- & -- & -- & -- & -- & --  \\
	 & 4 & 0.3915 & 49.1 & 23.43 & -18.80 & -18.69 & -18.36 & 0.09 & 0.08 & 0.06   \\
	 & 5  & 0.5797 & 55.4 & 23.87 & -20.04 & -19.52 & -19.29 & 0.27 & 0.17 & 0.14 \\\\
Q1436-0051  & limit$^7$ & 0.7377 & --  & -- & $>-19.09$ & $>-18.90$ & $>-18.03$ & $<0.11$ & $<0.09$ & $<0.04$ \\
		    & limit$^7$ & 0.9281 & -- & -- & $>-20.17$ & $>-19.90$ & $>-18.64$ & $<0.31$ & $<0.24$ & $<0.07$ \\
	& 2 & --  & 67.2, 72.2$^5$ & 24.99 & --  & -- & -- & -- & -- & --  \\
	& 5$^6$ & 0.9286 & 35.2 & 22.95 & -- & -22.90 & -21.80 & -- & 3.4 & 1.2  \\
	& 6$^{6}$ & 0.7401 & 45.1 & 21.62 & -- &  -23.10 & -22.30  & -- & 4.0 & 1.9   \\
	& 7$^{6}$ & 0.7387 & 53.9 & 22.44 & -- & -22.10 & -19.0 & -- & 1.6 & 0.8  \\
\hline
\end{tabular*}

$^1${Object IDs correspond to those in \citet{meiring11}.}
$^2${Impact parameters, given in kpc, are calculated from xy (arcsec) offsets taken from \citet{meiring11}. With the LDSS3 data presented in this work, we are able to determine accurate emission redshifts from which the actual impact parameter can be determined, in turn determining any association with respective absorption systems. }
$^3${Gunn r-band magnitudes for the targets are taken from \citet{meiring11}.}
$^4${Value taken from \citet{chun10}, based on K-band adaptive optics imaging.}
$^5${These impact parameters correspond to the absorber redshifts, $z=0.7377$ and $z=0.9281$ respectively. { Even though objects 5, 6, and 7 have been confirmed at these redshifts, it is possible that we are dealing with one or two galaxy group environments at $z\sim0.74$ and $z\sim0.93$ respectively, and therefore object 2 could still be at one of these redshifts.} The possibly perturbed morphology of object 5 suggests there may be a second unresolved interacting galaxy. }
$^6${These galaxies are confirmed as the host galaxies of strong neutral hydrogen absorber systems in this work. Objects 6 and 7 are at a similar redshift and likely interacting, so it is difficult to say which galaxy is causing the absorption at $z=0.7377$. We note, however, that object 6 is the closer in impact parameter, slightly stronger in flux, and has a slightly larger surface area. Photometric measurements are taken from \citet{meiring11}. }
$^7${Limits on i-band absolute magnitude and L$_{i}$/L$_{i}^{\ast}$ at the redshift of the absorption systems in each field for three spectral types. For Q1323-0021 and Q0826-2230, we do not detect the host galaxies of the absorption systems. Therefore, these galaxies must be below the limit for their respective fields. }
\end{minipage}
}
\end{table*}

\begin{landscape}
\begin{table}
\caption{Object spectral characteristics \label{tbl-characteristics}}
\begin{tabular}{llcccccccccccccccc}
\hline\hline
QSO & ID$^1$ & $\Delta$v  & $z_{em}$  & $f_{[O II]}$  & $f_{H\beta}$ & $f_{[O III]a}$ & $f_{[O III]b}$ & $f_{H\alpha}$  & SFR$_{[O II]}$ & SFR$_{H\alpha}$ & \multicolumn{2}{c}{12+log(O/H)} &  W$_{rest}^{X}$  & X \\ 
&  & (km s$^{-1}$) & & \multicolumn{5}{|c}{10$^{-17}$ ergs cm$^{-2}$ s$^{-1}$} & \multicolumn{2}{c}{M$_{\odot}$ yr$^{-1}$} & $l$ & $u$  & \AA  & \\ 
\hline
Q0826-2230 & 4  & -- &1.1457 & -- & -- & -- & -- & -- & $<0.40$ & -- & -- & -- & -- & -- \\
 	 & 5 & -- & 1.1457 & -- & -- & -- & -- & -- & $<0.40$ & -- & -- & -- & -- & -- \\
	  & 7 & -- & 0.1070 & 93.9 & -- & -- & -- & -- & -- & 32.4$^6$ & -- & -- & $<0.01^{4}$  & Ca II \\
	 & 7  & -- & 0.9512   & 93.9 & -- & -- & -- & -- & 57.3$^6$ & -- & -- & -- & $<0.02^{4}$ & Mg II  \\
Q1323-0021  & 2 & -- & --  & -- & -- & -- & -- & -- & -- & -- & -- & --  & $<0.50^{5}$ &  Mg II \\
	 & 3  & -- & -- & -- & --  & -- & -- & -- & $<0.55$ & --  & -- & -- & $<0.50^{5}$ &  Mg II \\
	 & 4  & -- & 0.3915 & -- & $<8.4$ & 4.68 & 18.03 & 16.31 & -- & 0.69 & -- & -- & $<0.44^{5}$ & Mg II   \\
	 & 5   & -- & 0.5797 & 78.48 & --  & -- & -- & -- & 15.1 & --  & -- & -- & $<0.24^{5}$ & Mg II \\
Q1436-0051  & 2  & -- & -- & -- & -- & -- & -- & --& $<2.5$ & -- & --  & --  & -- & -- \\
	& 5$^3$  & -78 & 0.9286 & 19.13 & -- & -- & -- & -- & 11.6 & -- & --& -- & $1.20\pm0.07$ & Mg II  \\
	& 6$^{2,3}$  &  -414 & 0.7401 & 72 & 115 & 43 & 46 & -- & 26 & -- & 7.15 & 9.03 & $1.14\pm0.08$ & Mg II  \\
	& 7$^{2,3}$ & -173 & 0.7387  & 64 & 77  & 29 & 31  & --  & 22  & -- & 7.25 & 9.01 & $1.14\pm0.08$ & Mg II \\
\hline
\end{tabular}

$^1${Object IDs correspond to those in \citet{meiring11}.}
$^2${Objects 6 and 7 are unresolved and therefore have been measured together. The text details the estimated flux ratio between the two, which has been used here to assign individual flux and SFR values.}
$^3${These galaxies are confirmed as the host galaxies of strong neutral hydrogen absorber systems in this work. Objects 6 and 7 are at a similar redshift and likely interacting, so it is difficult to say which galaxy is causing the absorption at $z=0.7377$. We have marked the galaxy closer in redshift (Object 7) as the host galaxy. We note, however, that Object 6 is the closer in impact parameter, slightly stronger in flux, and has a slightly larger surface area. }
$^4${Limits measured from the Magellan II Clay telescope MIKE spectra presented in \citet{meiring07}.}
$^5${Limits measured from the SDSS spectra for this QSO. We have assumed the redshift of the absorber ($z_{abs}=0.7160$) for objects 2 and 3 since no emission is detected for these objects in the LDSS3 spectra.}
$^6${Object 7 in the field of Q0826-2230 has a single emission line that could either be [O II] or H$\alpha$. Therefore, we have estimated the SFR for both lines at their corresponding redshifts. }
\end{table}
\end{landscape}

\section{Results}\label{section-results}

\subsection{Galaxy characteristics}

We present comparisons between each of the galaxy parameters found in our sample in conjunction with the literature for detected absorber host galaxies. Figure~\ref{fig-measurements} shows plots of the characteristics of our systems discussed in this section, including aggregate similar measurements on sub-DLAs and DLAs from the literature. These include the survey by \citet{peroux11a, peroux11b, peroux12} and references therein. The gray arrows indicate limits for the variables.  

Points marked with X's are from \citet{peroux11a,peroux12}, points marked with solid stars are from this work, and points marked with solid circles are from the literature. For the SFR plots, color indicates the emission line used for the estimate: blue for [O II], red for H$\alpha$, green for H$\beta$, purple for [O III], and black for Ly$\alpha$. New from this work are additional data on SFR and b as functions of absorber redshift, log N$_{\rm HI}$, and [Zn/H].

\subsubsection{SFR vs. logN$_{\rm HI}$, b, $z_{abs}$}

Including the data from \cite{peroux12}, 22 systems have measurements for both SFR and log N$_{\rm HI}$. We find no correlation in a Spearman rank order correlation test, resulting in $\rho=$0.023317 and P-value $=0.922269$. We also find no correlation between SFR and impact parameter. However, sample size for determined SFR in sub-DLAs and DLAs results in small number statistics.

\begin{figure*}
 
\includegraphics[width=1.0\textwidth]{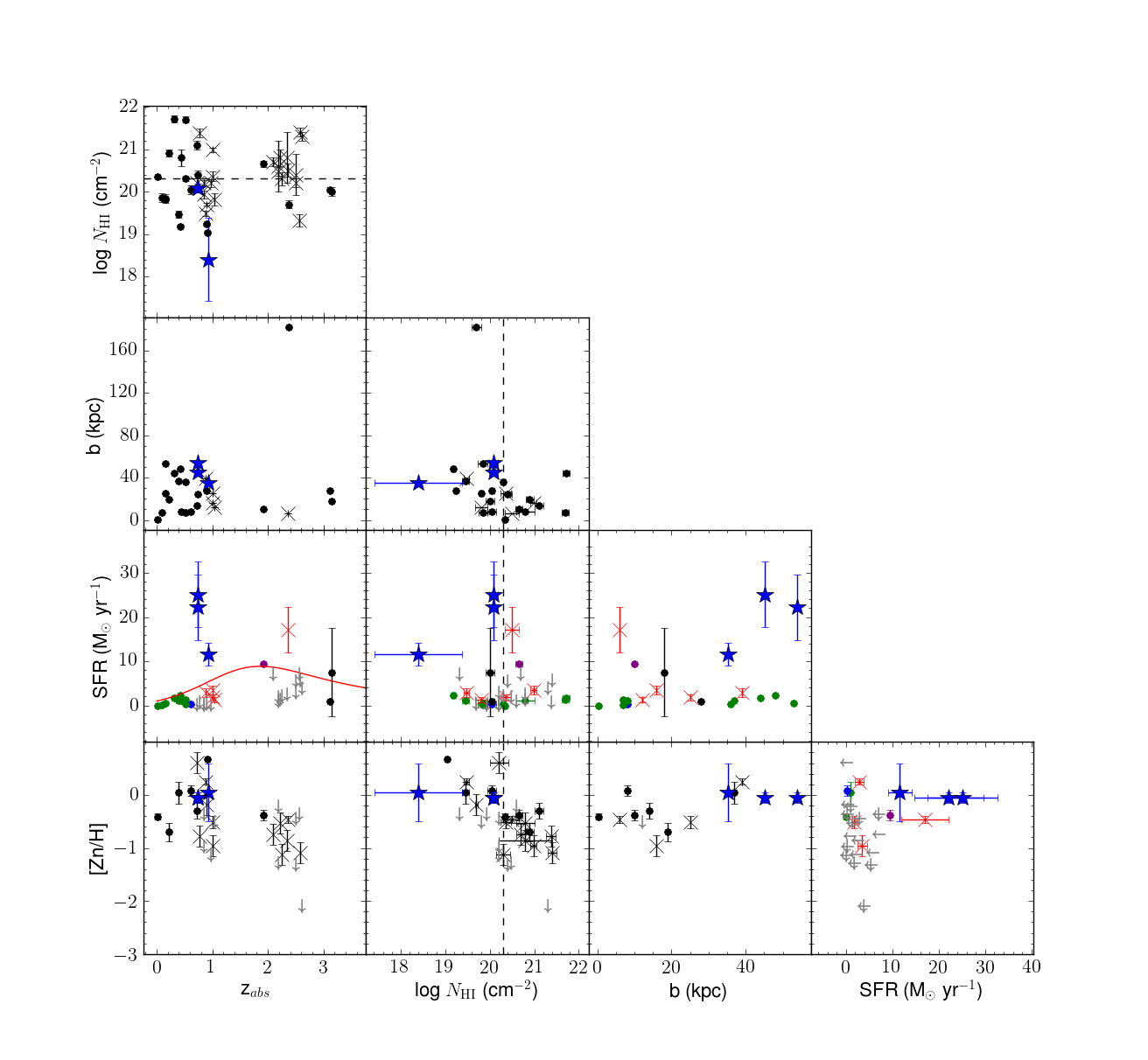}

\caption{Plots of various characteristics of sub-DLAs and DLAs in the literature. Blue star points mark the detections presented in this work. The X's mark detections from \citet{peroux11a, peroux12}. The points mark detections from previous studies in the literature. References: \citep{pettini00, lacey03, junkkarinen04, chen05, burbidge96, chun06, rao05, boisse98, lebrun97, gharanfoli07, rao06, steidel02, lanzetta97, lanzetta95, bergeron91, deharveng95, meiring07, cristiani87, bowen05, schulte05, bowen01, rosenberg06, steidel97, bergeron86, prochaska97, weatherly05, lu93, lu97, djorgovski96, christensen04, rao00, ellison05, pettini94, francis96, dodorico02, francis04, francis01, kulkarni05, cherinka09, noterdaeme09, fynbo10}. See Table 4 of \citet{peroux11a} for the full list of measurements. For the SFR plots, color indicates the emission from which the value was determined: blue for [O II], red for H$\alpha$, green for H$\beta$, purple for [O III], and black for Ly$\alpha$. SFRs have not been dust corrected. The solid red line in the SFR vs. $z_{abs}$ plot is an arbitrarily normalized Madau relationship from \citet{madau14}. Qualitatively, the trend in host galaxy SFR with absorber redshift is consistent with this trend. The blue outliers are the interacting pair in the field of Q1436-0051. The dashed vertical lines indicate the column density boundary between sub-DLAs and DLAs. The outlier in b at $\sim180$ kpc is a host galaxy that is part of a cluster environment reported in \citet{francis04b} and references therein. Gray arrows indicate limits for non-detections. The error bars on Q1436-0051 $z_{abs}=0.9281$ correspond to the range of possible HI column densities and Zn metallicities. See text for details. \label{fig-measurements}}

\end{figure*}

We find a significant correlation in a {Spearman rank order correlation test} between SFR and absorption redshift for detected galaxies, resulting in $\rho=0.692742$ and P-value $0.000710$. From the aggregate sample, including detections presented here, 20 systems have measurements for both of these parameters. {  Though the cosmic SFR density evolution is non-linear between $z=0$ and $z=4$, the data points in Figure~\ref{fig-measurements} are almost unanimously below $z\sim2$ where this evolutionary trend peaks. Compared to the cosmic SFR density relation observed in field galaxies by fitting an arbitrarily normalized Madau plot \citep[solid red line in Figure~\ref{fig-measurements},][]{madau14}, we see that our SFR data for sub-DLA and DLA host galaxies are consistent with the trend. The outliers depicted in blue with SFR higher than 15 M$_{\odot}$ yr$^{-1}$ are the two interacting galaxies in the field of Q1436-0051. The interaction is the likely cause of such high SFRs and is atypical for confirmed sub-DLA or DLA host galaxies.  }


\subsubsection{b vs. log N$_{\rm HI}$}

From the aggregate data, 27 systems have measurements for both impact parameter and log N$_{\rm HI}$. We find a moderate anti-correlation between these two parameters for detected galaxies, resulting in a Spearman $\rho=-0.462338$ and P-value$=0.015179$. This is consistent with the trend found by \cite{peroux12}, who report a correlation coefficient of $\rho =-0.56$ for a sample size of only five, and \citet[Figure 19]{rao11}, who report a strong 3$\sigma$ anti-correlation between the two variables. {  (We note that the Rao et al. sample includes many systems without redshift confirmations.) This includes the outlier at b$\sim180$ kpc. This object is part of a cluster at $z\sim2.38$, reported to be in a large cosmic filament by \citet{francis04}. The large impact parameter can be explained by the filamentary structure and group environment enhancing the radius of absorption. We find no change in the correlation by excluding this outlier. }

\subsubsection{[Zn/H] vs. b, SFR}


We find no correlation for the total sample between Zn absorption metallicity and impact parameter for detected galaxies. However, there appears to be the beginnings of a bimodality divided at b$\sim 30$ kpc. If we perform a Spearman rank order correlation test on the sample below 30 kpc, we again find no correlation. Testing the sample above 30 kpc also returns no correlation.  

While we find no correlation between Zn absorption metallicity and impact parameter, we can nevertheless estimate the metallicity gradient of our galaxies given an emission line metallicity (12+log(O/H)). Follow-up observations are necessary to obtain the emission lines necessary for this measurement, with the exception of Q1436-0051 objects 6 and 7 (presented above). We note that \citet{krogager12} and \citet{peroux12} find a positive correlation between metallicity and impact parameter ($r_{s}=0.70$). {  However, our selection of metal-rich systems is biased towards flat values of metallicity gradients. } 

We find no correlation between Zn metallicity and SFR for detected galaxies. \cite{peroux12} also do not find any correlation between these quantities and note that this may be due to significant dust in the most metal-rich systems (from which our sample is taken). 

\subsubsection{Galaxy Luminosity}

Assuming Sa type galaxies, we have looked for correlations between L$_{i}/$L$^{\ast}_{i}$, [Zn/H], SFR, and b. We find only very weak correlations between L$_{i}/$L$^{\ast}_{i}$ and [Zn/H] ($\rho=0.52$), as well as L$_{i}/$L$^{\ast}_{i}$ and SFR ($\rho=0.41$). These are supported by the expected metallicity-luminosity relationship for absorbers and the metallicity-luminosity relationship already observed in normal galaxies \citep{tremonti04}. 

All three of the detected galaxies that correspond to absorption systems (Q1436-0051 objects 5, 6, and 7) have L$>$L$^{\ast}$. These are clearly massive galaxies. The galaxies for which we have confirmed emission redshifts that do not correspond to absorption systems range from low mass to massive. Q0826-2230 has two massive galaxies with L$>$L$^{\ast}$ (but above the redshift of the QSO, and so we cannot confirm any corresponding absorption systems), while Q1323-0021 has two low-mass galaxies with L$<$0.2 L$^{\ast}$. The latter is a loose photometric definition of a dwarf galaxy. 

\subsection{Comparison with results in the literature}

\cite{peroux12} find SFR$<4$ M$_{\odot}$ yr$^{-1}$ for 3/3 detections at $z\sim1$ (1 sub-DLA and 2 DLAs) and 1/2 detections at $z\sim2$ (a sub-DLA). They find SFR$=17.1$  M$_{\odot}$ yr$^{-1}$ for 1/2 detections at $z\sim2$ (a DLA). They also report impact parameters of b$=6-39$ kpc. Further studies have also detected DLA and sub-DLA host galaxies in this same range, detecting them as close as 0.4 kpc and as far as 100 kpc \citep[e.g.][]{bowen05, chen05, rao11, peroux12}. \cite{zwaan05} report that the median impact parameter for a DLA host galaxy is about 7.8 kpc at $z\sim0$, and that at $z=0.5$($z=1.0$) 37\% (48\%) of DLA host galaxies will have impact parameters less than 1\arcsec.  For our sample in combination with \cite{peroux12} at $z\sim1$ (7 galaxies of which 2 are DLAs and 5 are sub-DLAs), we find an average impact parameter of 32 kpc with a median of 35 kpc, which is much higher than predicted by \cite{zwaan05}. This excess projected distance is likely due to our selection of only metal-rich absorption systems, compared to the typically metal-poor nature of DLAs. In addition, our detections are of one sub-DLA and one LLS, which may adhere to different projected distance trends than DLAs. 

The anti-correlation between impact parameter and log N$_{\rm HI}$ is consistent with the model predictions of \citet{yajima12}, which indicate that DLAs with lower gas density arise at larger impact parameters. \citet{noterdaeme14} study $\sim$100 strong DLA systems ($N_{\rm HI}$ $> 0.5\times 10^{22}$) and find that their attributes imply they arise in galaxies at very small impact parameters to the QSO line of sight ($<8$ kpc, though the host galaxies are not confirmed in their study). This also supports the anti-correlation {  between impact parameter and column density} found in confirmed host galaxies. 

Even considering the projected distance, we find an average impact parameter of $\sim4.3$\arcsec {  ($\sim 45$ kpc)}. Other studies have found DLA host galaxies at impact parameters $>20$ kpc, including \citet{bouche13} and \citet{kashikawa14}. Clearly the difficulty in detecting sub-DLA and DLA host galaxies is not solely due to the proximity to the QSO, but must include other factors such as dust and low surface brightness. Previous studies have suggested that the large cross-section of gas around massive galaxies is responsible for the higher metallicities found in sub-DLAs \citep{khare07, kulkarni10}, and that these systems therefore probe massive galaxies. The impact parameters and metallicities presented here support these claims.

{ We consider the potential difference between DLA and sub-DLA host galaxies in terms of metallicity gradients. \citet{chen05} present a sample of DLA host galaxies and their measured metallicity gradients in a comparable way and find significant decrements for  3 DLA host galaxies. They observe an abundance decrement by a factor of 3-5 from the centers of these host galaxies to 9-26 $h^{-1}$ kpc away. They note, however, that the R$_{23}$ uncertainties reduce the gradient for 2 of the 3 cases by 0.2-0.5 dex.  

If the trend is indeed that strong metallicity gradients exist for DLA host galaxies, there are multiple reasons for this not to be the case here. Previous studies with metallicity measurements typically have samples of isolated, unperturbed host galaxies with absorption metallicities in the sub-solar regime \cite[e.g.][who also find a mixture of metallicity gradients and non-gradients for both DLA and sub-DLA host galaxies, for a mixture of both metal-rich and metal-poor systems]{peroux12}. In contrast, objects 6 and 7 in the field of Q1436-0051 in this study are an interacting pair hosting a metal-rich (solar metallicity) sub-DLA. Each of these properties likely contribute to the lack of gradient in this system, and may indicate that this system is not representative of the current confirmed DLA and sub-DLA host galaxy population at large.}

\citet{fynbo13} report a positive detection of a metal-rich DLA host galaxy ($z_{abs}=2.583$, [M/H]$=0.0$) at an impact parameter of 16.2 kpc, which they claim is large for a DLA. They suggest that such large impact parameters of solar and super-solar metallicity DLAs could be due to processes other than outflows, such as tidal stripping similar to that of the Magellanic stream \citep{misawa09}. However, evidence for nearby galaxies that would cause this tidal stripping is not obvious. This scenario is most probable in the case of Q1436-0051 objects 6 and 7, which are interacting (see Figure~\ref{fig-q1436_soar}). These objects have high combined SFR (and from estimates, high individual SFR), indicating they may have the means to support strong metal-rich outflows. Therefore, the origin of this metal-rich system is ambiguous and could be stripped material or strong outflows. 

{ \citet{nestor11} report that the typical SFR$^{\ast}$ for galaxies at $z\sim0.7$ is between 1.6 and 9.3 M$_{\odot}$ yr$^{-1}$, making objects 6 and 7 in the field of Q1436-0051 rare, but still contribute significantly to the star formation density at this redshift. } Compared to other sub-DLA host galaxies, the two detections in our sample have much higher SFR. This may explain their high metallicity nature, all of them being super-solar and among the most metal-rich sub-DLAs known, again indicating a metallicity-luminosity relationship \citep{moller04, ledoux06, fynbo08, moller13}. In this scenario, the most metal-rich DLAs have the most luminous galaxy counterparts and are therefore easiest to detect. These galaxies would produce such metals through higher SFR, similar to typical emission-selected galaxies \citep[e.g.][]{fynbo99, rauch08}. 
Galactic outflows powered by supernovae feedback and star formation activity are expected to play a large role in the connection between DLAs and their host galaxies, explaining the metal-rich environments of DLAs observed up to 100 kpc from their host galaxies.

\citet{zwaan05} also find that at $z\sim0$ the most luminous galaxies are most likely to be associated with DLAs at large impact parameters. Our detections presented here are $z\sim1$, super-solar in metallicity, high impact parameter (at $b>30$ kpc), and are also more luminous than $L^{\ast}$ according to \citet{meiring11}. One of our detections is a sub-DLA (log N$_{\rm HI}$ $=20.08\pm0.11$). This places these host galaxies in the minority of DLA and sub-DLA host galaxies (fewer than 13\% have these characteristics). The majority of DLA host galaxies ($\sim$87\%) are expected to be sub-L$^{\ast}$ galaxies \citep{zwaan05}.  

\subsection{Effects of galaxy environment on absorption\label{section-origin}}

{The increased metallicity of the sub-DLA in Q1436-0051 and its lack of gradient may be explained by the galaxy interaction. Interactions strip gas and induce star formation, which in turn enriches the gas. This may occur in satellite galaxies in the host halo or manifest as wind-driven enriched gas. 

While in this study we select our targets based on strong H I absorption EW, there are numerous studies on Mg II-selected absorption systems that have associated these strong absorption features with environmental effects such as interactions. \citet{nestor06} have associated ultra-strong Mg II absorbers with similarly bright galaxies and environmental effects, finding an excess of $L>3L^{\ast}$ galaxies in the QSO fields in their study. \citet{gauthier13} also detected possible stripped gas in a galaxy group containing ultra-strong Mg II absorption. 

Additionally, \citet{kacprzak07} report a 3.2$\sigma$ correlation between disturbed galaxy morphologies and Mg II absorption equivalent width (for W$_{\lambda2796} < 1.4$ \AA). This correlation suggests that the strength of the perturbation in the galaxy determines the distance to which absorption can be detected. That is, the stronger the perturbation, the greater the impact parameter for absorbers of similar strengths. This could explain the apparently strongly interacting pair presented in this work, which has moderately strong Mg II absorption detected at $\sim 50$ kpc, compared to similar strength absorbers presented in \citet{kacprzak07} detected out to 80 kpc. The data in \citet{kacprzak07} also show a clear paucity of galaxies with minimal perturbation in the Mg II selected sample, indicating that Mg II absorbers may more frequently be found in morphologically perturbed galaxies. 

\citet{kacprzak10} study galaxy groups along the line of sight to Q1127-145. They find two galaxies at $z\sim0.328$ with $b=76.9$ kpc and $91.4$ kpc respectively that exhibit weak Mg II absorption ($W_{\lambda2796}^{r}=0.029$\AA). These two galaxies show unperturbed morphologies, supporting the theory that weaker absorption corresponds to unperturbed galaxies. They also confirm two new galaxies in a group of five total galaxies at $z\sim0.313$, four of which are within 100 kpc and one at $\sim240$ kpc. This galaxy group corresponds to moderately strong Mg II absorption and a DLA, with $W_{\lambda2796}^{r}=1.773$\AA. Each of the galaxies is considered to have undergone some type of interaction in its history, with two of them showing strongly disturbed morphologies and tidal tails. 

While previous studies such as \citet{gauthier13} and  \citet{nestor11} concentrate on ultra-strong Mg II absorbers (defined as having EW$>3$\AA), the sample presented in this work does not fall in this regime. Presented here are moderately strong Mg II absorbers with EW$\sim1$\AA chosen for their strong H I absorption EW and solar or super-solar metallicity, which overlap in part with the studies by \citet{nestor06} and \citet{kacprzak07}. We do not detect the presence of luminous infrared galaxies (LIRGs) in our study, which are typically associated with strong Mg II absorbers and starburst-driven outflows. Lyman-$\alpha$ emission selected galaxies are often dominated by mergers as well, but the galaxies here require UV spectra to determine their Lyman-$\alpha$ content. 

We note that there is evidence of galaxy environment playing a role in the absorption of higher ions as well, such as CIV and OVI \citep{chen09, johnson15}. Evidence of these higher ion absorption features have not been observed in the spectra presented here. OVI is only observable at $z>2.2$ in these MIKE spectra, while CIV is only observable at $z>1.2$ (while our confirmed host galaxies are at $z<1$). We also note that many of the above mentioned literature sources focus on Mg II absorption systems that are not necessarily DLAs or sub-DLAs, which we present in this work.

}

\subsection{Origin of absorbing gas}

{At low redshift, relatively high star-formation rates as those of the galaxies identified with the absorbers in the present paper are characteristic of interactions. \citet{bouche06}, \citet{bond01}, and \citet{prochter06} suggest that strong Mg II absorbers arise in galaxies with elevated star formation. This elevated star formation can be caused by galaxy interaction, driving gas into the halo to larger galactocentric distances via winds.  

The velocity dispersion for the absorbing gas for Mg II in the field of Q1436-0051 at $z=0.7377$ is $\Delta v\sim71$ km s$^{-1}$ according to \citet{meiring09b}, which is perhaps low for starburst-driven outflows at 50 kpc, which we might expect to have velocity spreads of many hundreds of km s$^{-1}$ as in \citet{nestor11}. Additionally, the kinematic structure of the absorption system is not complex \citep[as seen in][]{meiring08}, which is unusual for interacting galaxies \cite[e.g.][]{petitjean02}. As noted in \citet{meiring11}, the inclination of the two galaxy disks likely plays a large role in quiet velocity structure seen here. However, it has been shown in previous studies \citep[e.g.][]{bouche12} that galaxy alignment with the QSO along the galaxy minor axis is indicative of {\it inflows}, not metal enriched outflows. This, in combination with the simple kinematic structure of the absorption system, indicates that the absorption gas may not trace outflows at all. Instead, there may be a third undetected galaxy in this group responsible for the metal-rich absorbing gas. 

Tidally stripped gas is expected to be of a similar metallicity as the host galaxy, as is the case with objects 6 and 7 . Object 6 shows asymmetry in the direction of the QSO, indicating there may be a tidal tail extending in this direction containing the absorbing gas. This may even be another unresolved galaxy in the group, which would further complicate the kinematics.

Based on the asymmetric morphology and interaction of the galaxies, the highly elevated SFR ($>20$ M$_{\odot}$ yr$^{-1}$, possibly even $>40$ M$_{\odot}$ yr$^{-1}$), and the solar metallicity of the absorbing gas with no metallicity gradient, we conclude that there are several possible scenarios for the origin of this gas: starburst-driven winds expelling enriched material from the galaxies, tidally stripped enriched material (or a combination of these two), or a third satellite galaxy within the PSF of the QSO. Further high resolution observations are necessary to differentiate between these scenarios. 
 
}

 
\subsection{Galaxies with no absorption counterparts}

In addition to the two detections of $\rm HI$ absorption host galaxies, we have detected five galaxies that do not correspond to such absorption systems. Three of these galaxies, however, are likely above the emission redshift of the QSO. These detections are based on a single emission line detection. One of these could be below the redshift of the QSO, while the other two must be [O II] based on the non-detection of other lines that must be present in the spectra if it were H$\beta$, [O III], or H$\alpha$. 

For the other two galaxies with confirmed redshifts below that of the QSO, we have looked at the QSO spectrum for each of these fields and determined that within the wavelength coverage, there is no corresponding Mg II absorption (or other metal ions). This in and of itself is interesting, as each of these galaxies falls within $60$ kpc of the QSO sightline and other galaxies at similar impact parameters show strong absorption. 

As discussed in the individual objects sections, we place EW limits on absorption at each of these redshifts in the MIKE spectra for each QSO (Spectra provided by Joseph Meiring, private communication. Original spectra and measurements can be found in \citet{meiring08, meiring09b}). For the field of Q0826-2230, object 7 could be at $z_{em}=0.1070$ (H$\alpha$) or $z_{em}=0.9512$ ([O II]). In the archival MIKE echelle spectrum for this QSO, we place a 3$\sigma$ limit on Mg II $\lambda2796$ at $z=0.9512$ of EW$<0.02$ \AA. For $z=0.1070$, we place a 3$\sigma$ Ca II $\lambda3934$ limit of EW$<0.01$ \AA. The high resolution of MIKE (R$\sim28,000$) allows us to put very strict limits on these absorption features. No absorption is indicated at either of these redshifts for these species which, when detected, are among the most prominent absorption features. Objects 4 and 5 are at $z_{em}=1.1457$, which is above the QSO redshift. However, the QSO redshift is a lower limit for this field ($>0.911$), and so it is possible these objects are within the QSO redshift and exhibit no absorption. 

The field of Q1323-0021 has two galaxies confirmed at $z_{em}=0.3915$ and $z_{em}=0.5797$. The SDSS specrum of this QSO covers the wavelength range of Mg II for these redshifts. No absorption is detected. We place 3$\sigma$ Mg II EW of $<0.44$~ \AA ~and $<0.24$~ \AA, respectively. Both of these galaxies are sub-L$^{\ast}$ galaxies, which indicates that they may not show absorption due to a low absorption cross section. 

These absorption non-detections could be a result of the strong dependence of absorption on galaxy orientation with the QSO line of sight as presented by \citet{bordoloi11} and \citet{bouche12b}. \citet{bouche12b} show that Mg II absorption occurs in galaxies oriented along the major or minor axis towards the QSO, but no galaxies with Mg II absorption are found with inclination angles ($|\alpha|$) between $\sim20$ and $\sim60$ degrees. High resolution imaging is required in order to reveal the orientation and morphology of the galaxies in our sample and determine if galaxy orientation has played a role in the lack of detected absorption. 

{ Alternatively, the lack of Mg II absorption could be related to minimal or no perturbation to these galaxies, a correlation described in Section~\ref{section-origin}. It is not uncommon for studies to find galaxies with no Mg II absorption in fields of QSOs. \citet{kacprzak10} identify five galaxies along the line of sight to Q1127-145 with no absorption. All five galaxies ($z\sim0.3$) are at impact parameters $115<b<265$ kpc with $L<0.3L^{\ast}$, which is consistent with Mg II absorption occuring in luminous galaxies within ~100 kpc impact parameter, but not beyond. On the other hand, the simulations by \citet{kacprzak10} show that DLAs can occur even at ~200 kpc impact parameters. }

\section{Conclusions}\label{section-conclusions}

We present optical multi-object spectral data from LDSS3 on the Magellan II Clay Telescope of three QSO sightlines containing metal-rich, high $N_{\rm HI}$ absorption systems. We confirm in emission the host galaxies of a sub-DLA and a strong $\rm HI$ absorber with $17.45<$log $N_{\rm HI}<19.4$, which is 50\% of the sample herein.  We also confirm in emission five other galaxies within the fields that do not correspond to absorption systems within our wavelength coverage (three of which are likely above the QSO redshift based on our single line detection, and so cannot be observed in absorption). The hosts of two other DLAs remain undetected to i-band absolute magnitude limits of -18.78 and -20.36 at the redshift of the absorbers (Q1323-0021: $z_{abs}=0.7160$ and Q0826-2230: $z_{abs}=0.9110$, respectively). However, it should be noted that an early-type galaxy close the QSO line of sight in the field of Q1323-0021 is very likely the host galaxy, but does not yet have a successful isolated spectrum even after multiple studies have attempted it. 

For our detected host galaxies we are able to measure the SFR, impact parameter with the QSO line of sight, and velocity shift between the absorption and emission redshifts.  We report high mean SFR based on [O II] emission (SFR$ > 10$ M$_{\odot}$ yr$^{-1}$) for both of our absorber host galaxy detections. These are much higher than previous studies have found for sub-DLAs and DLAs, but are not incongruous given the super-solar metallicity of our targets.  We find the impact parameters of the QSO sightlines from the centers of the host galaxies to be in the range $b=35.2 - 53.9$ kpc, well within the ranges reported in the literature, which extend out to 100 kpc. These are, however, still higher than average for DLAs and sub-DLAs which are largely found within 35 kpc of the QSO line of sight. In addition, we have determined the L$_{i}/$L$^{\ast}_{i}$ values for each of the detected galaxies and find that those corresponding to absorption systems have L$>$L$^{\ast}$, making these host galaxies above average in all characteristics for DLAs and sub-DLAs. Meanwhile, the two galaxies at $z_{em}<z_{QSO}$ that do not show absorption have L$<0.2$L$^{\ast}$. These detections of high metallicity ([Zn/H]$=-0.05$), high luminosity host galaxies are consistent with a metallicity-luminosity relationship that has been suggested for DLAs and sub-DLAs. 

For object 5 in the field of Q1436-0051, we re-analyze the column density of the absorber at $z=0.9281$ in order to better understand its limits. From this, we report that the Zn metallicity of the system is very poorly constrained and is likely not super-solar as previously thought based on ionization corrections. We nonetheless detect the host galaxy of this system, and with followup study of the absorption system the characteristics may be very well constrained. 

For objects 6 and 7 hosting the sub-DLA in the field of Q1436-0051, we are able to measure the emission line metallicity and estimate the metallicity gradients at play in the system. We find no gradient, measuring $-0.01\pm0.006$ dex kpc$^{-1}$, which is consistent with that found for sub-DLAs and DLAs at similar impact parameters \citep{peroux12}. { We consider the conditions that create solar metallicity gas absorption at 35-54 kpc and conclude that there are several possible scenarios: the perturbation in the galaxies caused by their interactions produces highly increased star formation and star formation driven winds, tidally stripped metal-rich material from the two interacting galaxies, or a third undetected galaxy within the PSF of the QSO. The metallicity of the absorbing gas supports the first two scenarios (or a combination of the two), while the previously studied simple kinematic structure of the absorption system and apparent orientation of the galaxies indicates the absorbing gas may not be tracing outflowing material. Further observations are necessary to differentiate between these scenarios.}

We confirm the correlation between SFR and absorber redshift for detected host galaxies by combining our results with results from the literature at $z<2$ and show that it is qualitatively consistent with the cosmic star formation history, as depicted by the Madau plot. The data also support the anti-correlation in the literature between impact parameter and neutral hydrogen column density.

We show that our results are comparable to those in the literature for other DLA, sub-DLA, and Mg II selected samples, supporting the theory that environment plays a critical role in absorption kinematics and metallicity. Our results provide an important contribution to the small pool of previously confirmed host galaxies, which to date has numbered fewer than two dozen in a sample of over 12,000 DLA and sub-DLA absorption line systems \citep{noterdaeme12c}. We anticipate the onset of the integral field unit survey age to provide the number densities necessary to form a more complete picture of absorption systems and their host galaxies. 

\section*{Acknowledgements}

We thank the anonymous referee for the constructive comments and suggestions that improved this paper. This study includes data gathered with the 6.5 meter Magellan Telescopes located at Las Campanas Observatory, Chile. VPK acknowledges partial support from the National Science Foundation grant AST/1108830 (PI: Kulkarni). DGY acknowledges partial support from the John Templeton Foundation grant 12711 (PI: York). DGY and LAS acknowledge partial support from the John Templeton Foundation grant 37426 (PI: York). LAS acknowledges support from ERC Grant agreement 278594-GasAroundGalaxies (PI: Schaye). This research has made use of NASA's Astrophysics Data System. This research has also made use of the NASA/IPAC Extragalactic Database (NED) which is operated by the Jet Propulsion Laboratory, California Institute of Technology, under contract with the National Aeronautics and Space Administration.

\bibliographystyle{mn2e}
\bibliography{/data2/straka/Dropbox/bibliography/bibliography}

\bsp

\label{lastpage}

\end{document}